\documentclass[preprint2]{emulateapj}
\usepackage{natbib,graphicx,amsmath,longtable,amssymb,CJK,todonotes}

\def\lya{Ly$\alpha$~}
\def\nhi{\mbox{$N_{\rm HI}$}}

\def\LBstar{\mbox{$L_{\rm B}^{\star}$}}
\def\h100{\mbox{$h^{\rm -1}$}}
\def\kms{\mbox{$\rm ~km~s^{-1}$~}}
\def\kmso{\mbox{$\rm ~km~s^{-1}$}}
\def\cm-2{\mbox{$\rm ~cm^{-2}$}}
\def\mA{\mbox{$\rm m\AA$}}

\def\fcover{$f_{\rm cover}$~} 
\def\fcovero{$f_{\rm cover}$} 
\def\rvir{$R_{\rm vir}$}

\shorttitle{The Influence of Environment on the CGM}
\shortauthors{Yoon \& Putman}
\slugcomment{Accepted by ApJL on 7/7/2013}

\begin{document}
\begin{CJK*}{UTF8}{}

\title{THE INFLUENCE OF ENVIRONMENT ON THE CIRCUMGALACTIC MEDIUM}

\author{JOO HEON YOON (\CJKfamily{mj}윤주헌)$^{\star}$ and MARY E. PUTMAN}
\altaffiltext{}{Department of Astronomy, Columbia University, New York, NY 10027, USA;}
\email{$^{\star}$jhyoon@astro.columbia.edu}

\begin{abstract}
The effect of environment on the circumgalactic medium (CGM) is investigated through a comparison of \lya absorption line data in the Virgo Cluster and the field. This Letter uses the first systematic survey of background QSOs in and around the Virgo Cluster and large existing surveys of galaxies at low redshift. While previous studies found denser gas (higher equivalent width) closer to a galaxy (lower impact parameter), this correlation disappears in the Virgo environment. In addition, the covering fraction of the CGM is lower in the cluster environment than in the circumcluster environment and field. The results indicate that the CGM is suppressed for cluster galaxies while galaxies in the circumcluster environment have abundant CGM.  The truncation of the CGM may result in the quenching of star formation through starvation.  Our results also show that CGM surveys must consider the role of environment.

\end{abstract}

\section{Introduction}

Understanding the physics of gas provides a fundamental base for galaxy formation and evolution theories, since gas is essential to the formation of stars and galaxies in the universe.  Diffuse gas is a major fraction of the baryons in the universe and predominantly exists in the intergalactic medium \citep[IGM,][]{Fukugita1998a,Cen1999a}. Diffuse gas around galaxies, i.e., the circumgalactic medium (CGM), is the link between the IGM and galaxies.  
QSO absorption lines are one of the most promising ways of detecting the low column density CGM ($\nhi \sim 10^{13}\cm-2$).
There are numerous QSO absorption line studies that relate absorbers to either the CGM or IGM.   Galaxies are often found near \lya absorbers and there is an anti-correlation between the \lya equivalent width and impact parameter that indicates that the absorbers trace halo gas \citep[e.g.,][]{Chen2001a}.  However, not all of the absorbers seem to be directly related to individual galaxies, but rather can only be associated with the overall large scale structure  \citep[e.g.,][]{Tripp1998a,Chen2009a,Prochaska2011a}.

Galaxy clusters contain a large number of galaxies. The gas properties of the cluster galaxies are influenced by the hot intracluster medium (ICM) and the density of galaxies \citep{Byrd1990a,Moore1996a,Chung2007a}. The CGM of galaxies in a cluster environment has largely not been investigated. We completed a survey of background QSOs toward the environment of the Virgo Cluster and found 43 \lya absorbers \citep[][hereafter Y12]{Yoon2012a}. This study, together with the well-defined galaxy catalogs in this region \citep{Binggeli1985a} and spectroscopic survey data \citep[Sloan Digital Sky Survey, SDSS;][]{York2000a}, enables us to now examine the CGM in a cluster environment. This Letter examines the properties and association of \lya absorbers with galaxies in and around the Virgo Cluster.  We probe the CGM of galaxies in {\it cluster, circumcluster,} and {\it field environments} and discuss the physical mechanisms influencing the CGM.

\section{Data}
\label{data.sec}

This section describes the absorber and galaxy data used for this Letter. In and around the Virgo Cluster, 43 \lya absorbers toward 23 background QSOs were found in the velocity range $700<v<3000\kmso$ as described in Y12. The galaxies are selected within the volume of 2.5\rvir~in projection of the Virgo Cluster (\rvir~$=1.57$ Mpc centered on  M87) and the velocity range $700<v<3000\kmso$.  The full velocity range of the Virgo Cluster galaxies extends to $\sim -700\kms$\citep{Binggeli1993a}, but our \lya absorbers cannot probe the entire velocity range due to the Milky Way damping wing (see Y12). Our coverage from 700-3000\kms contains at least 84\% of the galaxies in the Virgo region, so we are able to examine the effect of environment on the CGM for a majority of the galaxies.
Throughout this Letter, we define a galaxy as in a {\it cluster environment} if it is within \rvir~in both projection and line-of-sight distance. The remaining galaxies within 2.5\rvir~in projection are defined as a {\it circumcluster environment}. When we consider galaxies in both the cluster and circumcluster environment, we refer to it as the {\it Virgo environment}.

For our Virgo galaxy catalog, we merge the SDSS DR7 spectroscopic catalog \citep{Abazajian2009a} with galaxies from the NASA/IPAC Extragalactic Database (NED) in the same spatial and velocity range. We selected galaxies brighter than $r < 17.77$ in the SDSS catalog \citep{Strauss2002a}.  For the galaxies from NED, we selected galaxies brighter than 17.88 mag in the Johnson $B$-band since this corresponds to $r=17.77$ when we perform a one-to-one comparison of magnitude for the galaxies observed in both $B$-band and $r$-band. The luminosity of the galaxies selected ranges from 0.0002\LBstar~to 2.5\LBstar. This is based on distances from the literature, the Virgo inflow model, and the Hubble flow (in that order) as provided by M. Haynes (2012, private communication) and described in \citet{Haynes2011a}. We also use these distances to define galaxies as within the virial radius or outside the virial radius of Virgo. Unfortunately, many distance estimates in this region of the sky are uncertain \citep[e.g.,][]{Cortes2008a}. As a check, we looked at the radial velocities to define cluster and circumcluster galaxies and found similar results.

We adopt two field samples to compare to the cluster environment.  For the first field sample, we select 81 low redshift ($v<6000\kmso$) \lya absorbers from the literature \citep{Bowen2002a,Penton2000a,Penton2004a,Sembach2004a,Richter2004a,Danforth2006a,Danforth2010a,Wakker2009a,Prochaska2011a} that are also covered by the SDSS, 2dF \citep{Colless2001a}, or 6dF surveys \citep{Jones2004a,Jones2009a}. We adopt 6000\kms as the upper velocity cut because using the 3000\kms cut used for our Virgo sample results in small number statistics.  To match the magnitude limit for SDSS galaxies ($r < 17.77$), we simply match SDSS $r$-band magnitudes and 2dF/6dF $b_{J}$ magnitudes of galaxies observed by both the SDSS and 2dF/6dF.  We apply $b_{J} < 18.46$ as the magnitude cut for the 2dF/6dF galaxies.  For the second field sample, we adopt the absorbers and galaxies from \citet[][hereafter P11]{Prochaska2011a}. This field sample is at a higher redshift, $z<0.2$, and the galaxies span a brighter luminosity range. We refer to this as the P11 sample throughout. The redshift uncertainties of the SDSS, 2dF, and 6dF surveys are $\sim 30\kms$\citep{Abazajian2003a}, $\sim 85\kms$\citep{Colless2001a}, and $46-116\kms$\citep{Jones2004a}, respectively. About 60\% of the field absorbers and galaxies are covered by SDSS.

The velocity intervals used to associate galaxies with \lya absorbers are that they are within $\pm$200 or $\pm$400\kms of each other. The more conservative cut, $\pm$200\kmso, is chosen as an analog to gas in the halo of a Milky Way-like galaxy, and $\pm$400\kms is set to be compared with the P11 sample. These velocity intervals must be interpreted with some caution in that the velocity structure in this dense environment is distorted.  This velocity distortion will cause more uncertainty in determining if a galaxy and an absorber are physically associated.

\section{Results}

We probe the CGM in various environments in this section by pairing \lya absorbers to the nearest galaxy. Pairing a galaxy and its absorber is not trivial because there are numerous galaxies and multiple QSO sightlines in the Virgo environment. We describe how we pair a galaxy and an absorber as follows. (1)  We find the nearest galaxy to a single sightline in projection. (2) If a \lya absorber is found toward the sightline within a velocity interval of the galaxy, the galaxy and the absorber are paired. This paired galaxy and absorber are excluded in this pairing algorithm afterward. (3) If a galaxy has two nearest absorbers with the same impact parameter, the one with a larger equivalent width is paired with the galaxy. (4) If the nearest galaxy has no \lya absorber within the chosen velocity interval, an upper limit for a \lya detection is set for that galaxy. This is then a galaxy-upper limit pair. Other galaxies in the same velocity interval as this galaxy-upper limit pair are then removed from being paired with that sightline. Processes (1) -- (4) are iterated until all galaxies within a limiting radius of 1 Mpc around sightlines are paired with either absorbers or upper limits. 
With this method, we found 42 (43) galaxy-\lya absorber pairs and 67 (33) galaxies do not show \lya absorbers within the velocity intervals $\Delta v=200(400)\kmso$.
 
  Below 1000\kmso, the sensitivity to an absorption line detection drops due to the Milky Way damping wing, as explained in Y12. This can cause the covering fraction, \fcovero, to be underestimated. In order to alleviate this, we compute upper limits for \lya detections as a function of wavelength for all spectra. These upper limits increase exponentially with proximity to the damping wing. We find a mean $W_{\rm ul}$ of all sightlines as a function of wavelength and use this mean for galaxies from 700 to 1000\kmso. Based on the $W (>100~\mA)$ distribution of our \lya absorbers, the probability, $P$, of an absorber to lie between $100~\mA<W<W_{\rm ul}$ is calculated. We put less weight, $1-P$, on that galaxy when computing \fcovero. This correction does not cause any significant changes in the results.

\subsection{$\rho-W$ Anti-correlation}

\begin{figure}
\begin{center}
  \includegraphics[width=\columnwidth]{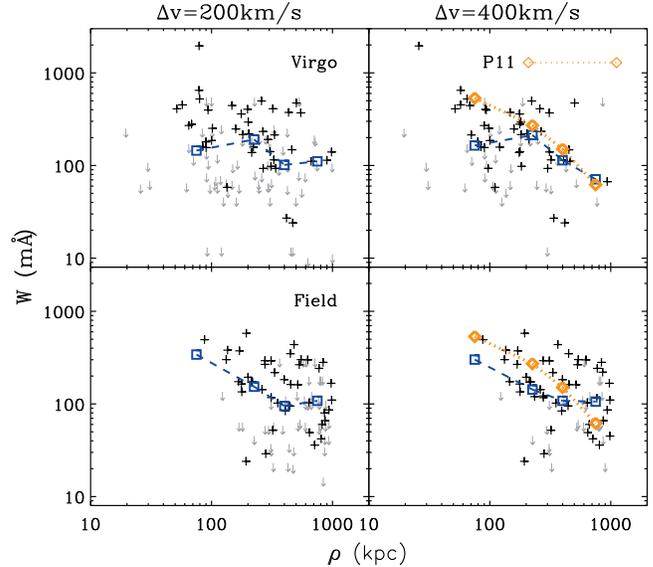}
\caption{Correlation between equivalent width ($W$) of \lya absorbers and impact parameter ($\rho$) within the velocity intervals of $\pm$200 and $\pm$400\kmso. Downward arrows present upper limits for non-detection of \lya absorbers. The pairs are selected in the same velocity interval in each column, as noted at the top. The top row shows the galaxies in the Virgo environment and the bottom row presents the field galaxies.  The median line (dashed line with squares) is illustrated in the bins of 0-150, 150-300, 300-500, 500-1000 kpc. The median for the P11 sample is shown by the dotted line with diamonds. }
\label{f1.fig}
\end{center}
\end{figure}

It is known that the line strength of \lya absorbers increases with decreasing distance to the nearest galaxies \citep[e.g.,][]{Chen2001a}. Here, we investigate the relationship between impact parameter, $\rho$ (distance from a galaxy to an absorber), and the equivalent width of an absorber, $W$, in the Virgo (cluster+circumcluster) and field environments. For both $\Delta v=200\kms$ and 400\kmso, the $\rho-W$ anti-correlation cannot be found in the Virgo environment (top row in Figure~\ref{f1.fig}). This is particularly due to the pairs with small impact parameters. We also paired gas-rich galaxies from the ALFALFA galaxy catalog \citep{Giovanelli2007a,Haynes2011a} with \lya absorbers in the Virgo environment. They also do not show the anti-correlation. Our field galaxy sample (bottom row of Figure~\ref{f1.fig}) does show the anti-correlation, decreasing median $W$ with $\rho$, but with a large scatter. The P11 galaxies (dotted lines with diamonds in the right panels) have a consistent slope with our field galaxies. We were not able to examine the $\rho-W$ anti-correlation in the separate cluster and circumcluster environment, or by a galaxy type due to low number statistics.

\subsection{The Covering Fraction}
 
\begin{figure}
\begin{center}
  \includegraphics[width=\columnwidth]{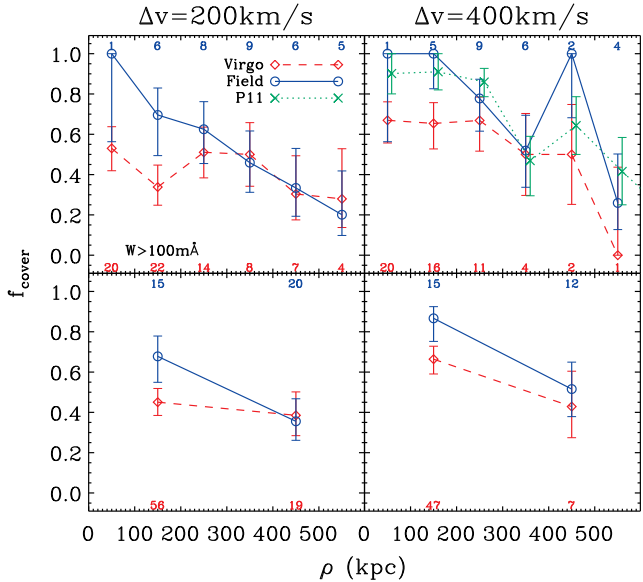}
\caption{Covering fraction of Virgo and field galaxies in bins of 100 (top) and 300 kpc (bottom). The velocity intervals $\pm$200 and 400\kms are used in the left and right columns, respectively. The numbers presented in each bin show the number of both galaxy-absorber and galaxy-upper limit pairs. The error bar indicates $1\sigma$ from likelihood distribution. The P11 field sample is shown in the upper right plot for comparison.} 
\label{f2.fig}
\end{center}
\end{figure}

\begin{figure}
\begin{center}
  \includegraphics[width=\columnwidth]{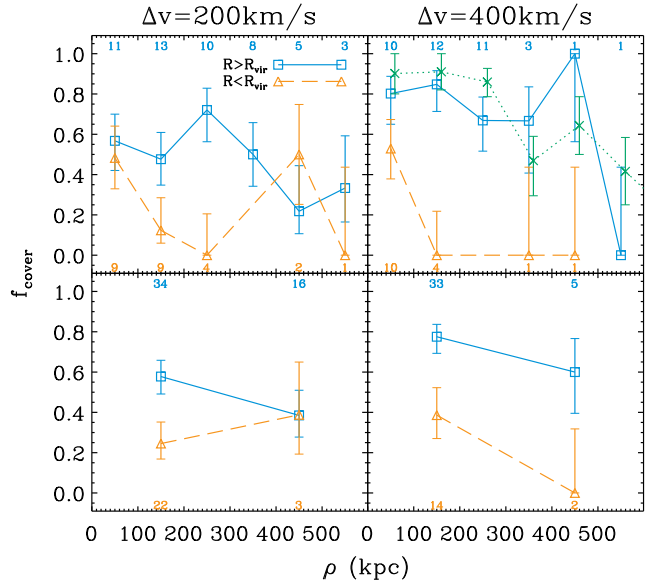}
\caption{Covering fraction of galaxies in the cluster and circumcluster environment is presented in bins of 100 (top) and 300 kpc (bottom). The velocity intervals, numbers, error bars, and P11 sample are illustrated as Figure~\ref{f2.fig}. } 
\label{f3.fig}
\end{center}
\end{figure}

In order to further investigate the CGM of galaxies, we compute the covering fraction, \fcovero, and examine it in various environments. The number ratio of galaxy-absorber pairs to all galaxy-absorber and galaxy-upper limit pairs is calculated as \fcover in bins of 100 kpc and 300 kpc for $W>100$ \mA~absorbers only with the sightlines that have a better sensitivity than 100~\mA~in Figures~\ref{f2.fig} and \ref{f3.fig}. We show an impact parameter range $0<\rho<600$ kpc because there are very few galaxy-absorber or galaxy-upper limit pairs at $\rho>600$ kpc (see top-left panel of Figure~\ref{f1.fig}).  The covering fraction is computed with the velocity intervals of $\pm$200 and $\pm$400\kmso.

 The covering fraction for the galaxies in the Virgo environment shows no clear decrement with increasing $\rho$ (dashed line with diamonds in Figure~\ref{f2.fig}), while that of the field galaxies shows an decreasing \fcover with $\rho$ (solid line with open circles in Figure~\ref{f2.fig}). The covering fraction of the field galaxies is in agreement with that of the P11 galaxies (dotted line with crosses in upper right panel of Figure~\ref{f2.fig}).  For better number statistics, we use a 300 kpc bin in the bottom row of Figure~\ref{f2.fig}. The covering fraction is lower for the Virgo  galaxies, \fcover$=0.45_{-0.07}^{+0.07}(0.66_{-0.07}^{+0.06})$, than for the field galaxies, \fcover$=0.68_{-0.13}^{+0.10}(0.87_{-0.11}^{+0.06})$, within 300 kpc with the velocity interval $\pm$200(400)\kmso.  The covering fractions outside 300 kpc are indistinguishable from each other, but at this distance, we are no longer tracing the CGM within a galaxy's virial radius. 
 
In Figure~\ref{f3.fig}, we separate the Virgo galaxies into those within \rvir~(cluster environment) and outside \rvir~(circumcluster environment; see Section~\ref{data.sec}).  As in Figure~\ref{f2.fig}, we examine the data in 100 and 300 kpc bins. The covering fraction for galaxies outside \rvir~with $\Delta v=400\kms$ is consistent with the P11 galaxies as shown in the upper right panel. This indicates that the difference seen in the previous plots are due to galaxies within \rvir. Indeed, predominantly using the 300 kpc bins (bottom row), the galaxies outside \rvir~(solid line with squares) show a higher \fcover$=0.58_{-0.09}^{+0.08}(0.78_{-0.08}^{+0.06})$ than those within \rvir~(long-dashed line with triangles), \fcover$=0.25_{-0.08}^{+0.11}(0.39_{-0.12}^{+0.14})$, within 300 kpc. 
With 100 kpc bins, there is a rise at 0-100 kpc for galaxies within \rvir~with both velocity intervals. We thoroughly investigated this rise and found that it is due to galaxies in Virgo's infalling substructures which are described in Y12 and references within. This is also the case for the increase in the $400<\rho<500$ kpc bin with $\Delta v=200\kmso$ (see Section~\ref{discussion.sec} for details).  These contaminating galaxies are few and therefore the effect of the substructures is lessened in the 300 kpc bin plot.  

\section{Discussion}
\label{discussion.sec}

Previous studies detected \lya absorption lines within $\sim300$ kpc of galaxies in almost all cases \citep{Chen2001a,Bowen2002a,Wakker2009a,Prochaska2011a}.  Although the galaxies in these studies include some group members \citep{Wakker2009a}, the vast majority of the galaxies are not in an environment as dense as the Virgo Cluster. 
We have shown that the CGM properties vary with the Virgo and field environments. The galaxies in the Virgo environment do not show the $\rho-W$ anti-correlation seen for the field and P11 galaxies. Also, the covering fraction of the CGM for the Virgo environment galaxies is lower than that of the field/P11 galaxies within 300 kpc.   P11 used the number of neighbors to claim that their derived CGM properties are independent of local environment; this does not seem to be the case when the cluster environment is probed.
Our results are consistent with \citet{Wakker2009a} who found a lower \fcover for group galaxies (0.57) than field galaxies (1.00) with $\Delta v=400\kmso$. Their covering fraction for group galaxies is between our cluster galaxies (0.39) and circumcluster galaxies (0.78).  Another recent study found the CGM is present around early-type galaxies \citep{Thom2012a}. We performed visual classification of our galaxy-upper limit pairs and find that early-type galaxies within \rvir~do not show any associated \lya absorbers. Therefore, our results indicate that the presence of the CGM around early-type galaxies is determined not by galaxy type, but by environment.  A dense environment is playing an important role in suppressing the CGM.  

Disk gas stripping by a cluster environment has been observed in many previous studies. For instance, \citet{Chung2007a} found gas being stripped from galaxy disks at $0.6-1$ Mpc in the Virgo Cluster (Virgo \rvir~$=1.57$ Mpc).  Since the CGM is more diffuse than disk gas, and more loosely bound, it is more vulnerable to stripping. Gas stripping by ram pressure will occur if $\rho_{\rm ICM} v_{\rm gal}^2>\Sigma_{\rm gas} v_{\rm rot}^2 R^{-1}$ \citep[][modified by \citealt{Vollmer2001a}]{Gunn1972a} where $\rho_{\rm ICM}$ is the ICM density, $v_{\rm gal}$ is the relative velocity of a galaxy to the ICM, $\Sigma_{\rm gas}$ is the column density of the galaxy's gas, $v_{\rm rot}$ is the rotation velocity of a galaxy, and $R$ is the radius of a galaxy. For the CGM, we assume the parameters as follows: $\Sigma_{\rm gas}=10^{15}(10^{14})\cm-2$, $v_{\rm rot}=200\kmso$, $R = 100 (300) $ kpc, and $v_{\rm gal}=1000\kms$ as a representative value for ram pressure stripping in the Virgo Cluster \citep{Vollmer2001a}. Then, the ICM density required to cause CGM stripping is $\rho_{\rm ICM} \sim 10^{-10} (4\times10^{-12})~\rm cm^{-3}$. This ICM density is many orders of magnitude lower than the representative density of the ICM for ram pressure in the Virgo Cluster, which is $\rho_{\rm ICM} = 10^{-4}~\rm cm^{-3}$ \citep{Vollmer2001a}. Therefore, ram pressure stripping of the CGM will occur earlier than disk gas stripping and in an environment that is not dense enough to remove disk gas. 

The truncation of the CGM for the cluster galaxies can be attributed to either the rapid integration of stripped gas into the ICM or the lack of the accretion of warm gas.  While fast integration of stripped gas to the ICM is investigated in theoretical studies \citep[e.g.,][]{Tonnesen2010a}, gas accretion onto a galaxy in different environments still remains to be probed.
In any case, the CGM truncation removes future fuel for star formation and will cause a galaxy to evolve onto the red sequence through a process called {\it starvation} \citep{Larson1980a}. Starvation is expected to begin around the cluster virial radius by stripping outer halo gas in simulations \citep{Tonnesen2007a}. Our results provide observational evidence for this starvation.

We do not see the effect of CGM stripping in the circumcluster environment as the covering fraction of the CGM for the circumcluster galaxies is in agreement with that of the field and P11 galaxies. Further studies should attempt to examine at what radii the CGM is stripped as a galaxy enters a cluster given the low densities required to strip the CGM.   This may be more straightforward to investigate in a more regular cluster than Virgo which has numerous infalling substructures. The CGM may be protected in the circumcluster environment by the fact that the galaxies are moving with the IGM into the cluster (see Y12). 

We see somewhat of an increase in \fcover within 100 kpc and $400<\rho<500$ kpc in the top left panel ($\Delta v=200\kmso$) of Figure~\ref{f3.fig} even for the galaxies within \rvir. The increase within 100 kpc is also seen in the $\Delta v=400\kms$ panel (right). Given the typical lack of absorbers within \rvir, it is interesting to examine the reason for these increases.  The uncertainty in the distance estimates, as explained in Section~\ref{data.sec}, may play somewhat of a role. For instance, though most things remain very similar, the rise at $400<\rho<500$ kpc disappears when velocities instead of distances are used to define cluster galaxies.  Most of the galaxy-\lya absorber pairs causing the \fcover increase can be linked back to Virgo substructures that are infalling or are thought to have fairly recently fallen in.  Cluster B is one of the main culprits.  Cluster B is within the virial radius of the cluster in projection and at the same distance as the main cluster \citep{Mei2007a}; however, it contains many gas-rich galaxies and is thought to be an in-fallen structure (see Y12 for more details). 
 One other particularly interesting galaxy-\lya absorber pair within \rvir~ is in the vicinity of the subcluster containing M86. The pair is $\sim200$ kpc in projection from M86 and not far in position-velocity space from a long tail of atomic hydrogen and H$\alpha$ emission that is connected to M86 \citep{Oosterloo2005a,Kenney2008a}.  The M86 subcluster may have recently fallen into Virgo from the backside and the tail is thought to be stripped gas residing within the cluster \citep{Oosterloo2005a}.

\section{Conclusion}

By examining galaxies and \lya absorbers in cluster, circumcluster, and field environments, we find the following.
\begin{enumerate}
\item The typical anti-correlation between impact parameter and a \lya equivalent width is not found for the Virgo galaxies and absorbers.
\item The CGM covering fraction for Virgo galaxies is lower than that of field galaxies.  In particular, galaxies within the virial radius of the cluster show a lower \fcovero, while galaxies in the circumcluster environment have a comparable \fcover to field galaxies.
\item  The results indicate the cluster environment suppresses the CGM, and that the presence of a warm CGM depends not on galaxy type, but on environment.  
\end{enumerate}

The CGM can be easily suppressed by gas stripping processes in the cluster environment (e.g., ram pressure); however, the prevention of gas accretion to individual galaxies may also play a role. This removal of the CGM will result in starvation, and quench star formation. Further studies of CGM variation with environment will provide additional insight into the link between the surrounding IGM, the CGM, and galaxies.

\acknowledgments
We thank Hsiao-Wen Chen for the extensive discussions on how to relate a galaxy to an absorber in a cluster environment, and Martha Haynes for generously providing distance information for the Virgo galaxies.  We also thank Jacqueline van Gorkom, J. Xavier Prochaska, David Schiminovich, and Joshua Peek for useful discussions.   We acknowledge funding from NASA grant HST-GO-11698, the Luce Foundation, and NSF CAREER grant AST-0904059.


\end{CJK*}
\end{document}